\documentclass[journal]{IEEEtran}

\usepackage{cite}
\usepackage[pdftex]{graphicx}
	\ifCLASSINFOpdf
	 	\usepackage[pdftex]{graphicx}
  	\else
	\fi
\usepackage[cmex10]{amsmath}
\usepackage{amsthm}
\usepackage{array}
\usepackage{amssymb}

\begin{document}

\title{Information Theoretic Exemplification of the Impact of Transmitter-Receiver Cognition on the Channel Capacity}
\author{Nima~S.~Anzabi-Nezhad,
        Ghosheh~Abed~Hodtani,~and~Mohammad~Molavi~Kakhki
\thanks{N. S. Anzabi-Nezhad is with the Department of Electrical Engineering, Ferdowsi University of Mashhad, Iran, email: nima.anzabi@gmail.com}
\thanks{G. A. Hodtani is with the Department of Electrical Engineering, Ferdowsi University of Mashhad, Iran, email: ghodtani@gmail.com}
\thanks{M. Molavi Kakhki is with the Department of Electrical Engineering, Ferdowsi University of Mashhad, Iran, email: molavi@um.ac.ir}}

\maketitle

\begin{abstract}
In this paper, we study, information theoretically, the impact of transmitter and or receiver cognition on the channel capacity. The cognition can be described by state information, dependent on the channel noise and or input. Specifically, as a new idea, we consider the receiver cognition as a state information dependent on the noise and we derive a capacity theorem based on the Gaussian version of the Cover-Chiang capacity theorem for two-sided state information channel. As intuitively expected, the receiver cognition increases the channel capacity and our theorem shows this increase quantitatively. Also, our capacity theorem includes the famous Costa theorem as its special cases.
\end{abstract}

\IEEEpeerreviewmaketitle

\begin{IEEEkeywords}
transmitter-receiver cognition, Gaussian channel capacity, correlated side information.
\end{IEEEkeywords}

\section{Introduction}
Information theoretic study of the impact of transmitter and or receiver cognition on the channel capacity is a new idea and an important research issue. For example one channel from view points of two receivers with different cognition and information on the channel, may have different capacities. The cognition at the transmitter or receiver can be described by the usual concept of information theory i.e., side information.

Side information channels have been extensively studied since the initiation by Shannon \cite{shannon} and the subsequent study by Kusnetsov-Tsybakov \cite{kusnetsov}. The capacity of channel with side information (CSI) known causally only at the transmitter and only at the receiver has been determined by Gel'fand-Pinsker(GP) \cite{GP} and Heegard-El Gamal \cite{HG} respectively. Considering the GP theorem for the Gaussian channel, Costa \cite{costa} obtained an interesting result, i.e., the channel capacity in the presence of interference known at the transmitter is the same as the case without interference. Having extended the above results, Cover-Chiang \cite{coverchiang} established a general capacity theorem for the channel with two-sided state information. We have many other important researches in the literature, e.g.\cite{sajafar,keshet2008,merhav2007}. The results obtained for side information point to point channel have been extended, at least at special cases, to multiuser channels \cite{sigurjonsson2005,kim2004,khosravi2010,philosof2009,steinberg20052}.

As mentioned above, our motivation was the fact that cognition of the transmitter and receiver can affect the channel capacity. In order to quantify this effect, we illustrate the cognition as state information dependent on the channel noise and or input. Then we derive a capacity theorem and prove that, as expected, the receiver cognition increases the channel capacity  and our theorem shows this increase quantitatively. Our capacity theorem, while revealing the importance of Costa theorem, is a more general theorem and includes the Costa theorem as special cases.\\
In the remainder of this section we briefly review the Cover-Chiang, the Gel'fand-Pinsker and the Costa theorems.

\begin{figure}[!t]
\centering
\includegraphics[width=3.5in]{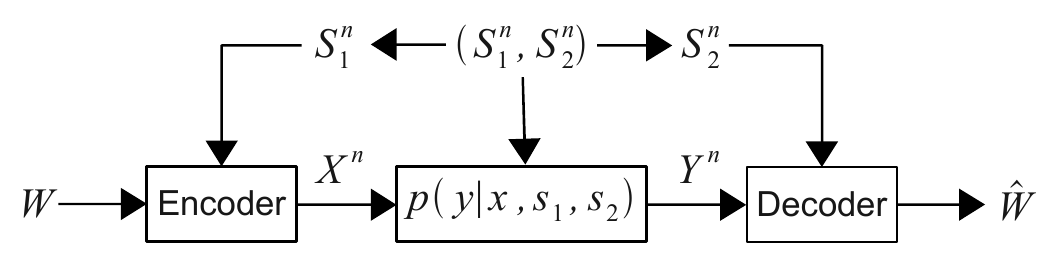}
\caption{Channel with side information available non-causally at the transmitter and at the receiver.}
\label{figure1}
\end{figure}

\textit{Cover-Chiang Theorem:}
Fig.\ref{figure1} shows a channel with side information known at the transmitter and at the receiver.
$ X^{n} $ and $ Y^{n} $ are the transmitted and received sequences respectively. The sequences $ S_{1}^{n} $ and $ S_{2}^{n} $ are the side information known non-causally at the transmitter and at the receiver respectively. The transition probability of the channel $ p\left( y\mid x,s_{1},s_{2}\right)  $ depends on the input $ X $, the side information $ S_{1} $ and $ S_{2} $. If the channel is memoryless and the sequences $ \left(S_{1}^{n},S_{2}^{n}\right)  $ are  independent and identically distributed (i.i.d.) random variables under $ p\left(s_{1},s_{2} \right)  $, then the capacity of the channel is \cite{coverchiang}:
\begin{equation}
C=\max_{p\left(u,x\mid s_{1} \right) }\left[I\left(U;S_{2},Y \right)-I\left( U;S_{1}\right)   \right] \label{eq.1} 
\end{equation}
where the maximum is over all distributions:
\begin{equation}
p\left(y,x,u,s_{1},s_{2}\right) =p\left(y\mid x,s_{1},s_{2} \right)p\left(u,x\mid s_{1} \right)p\left(s_{1},s_{2} \right) \label{eq.2}  
\end{equation}
and $ U $ is an auxiliary random variable for conveying the information of the known $ S_{1}^{n} $ into $ X^{n} $.

It is important to note that the Markov chain:
\begin{equation}
S_{2}\longrightarrow S_{1}\longrightarrow UX \label{eq.3} 
\end{equation}
is satisfied for all above distributions. 

\begin{figure}[!t]
\centering
\includegraphics[width=3.5in]{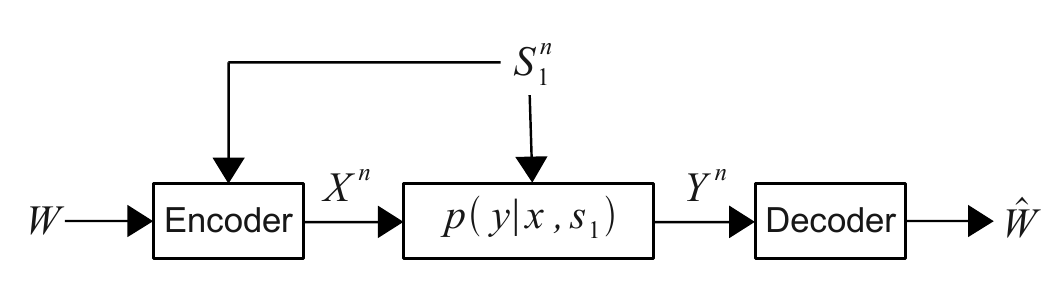}
\caption{Channel with side information known at the transmitter.}
\label{figure2}
\end{figure}

\textit{Gel'fand-Pinsker Theorem:}
The situation $ S_{2}=\phi $ (no side information at the receiver) leads to the Gel'fand-Pinsker theorem \cite{GP}:The memoryless channel with transition probability $ p\left(y \mid x,s_{1} \right)  $ and the side information sequence $ S_{1}^{n} $ (which is i.i.d. $ \sim p\left(s_{1} \right)  $) known non-causally at the transmitter (Fig.\ref{figure2}) has the capacity
\begin{equation}
C=\max_{p\left(u,x \mid s_{1} \right) } \left[I\left(U;Y \right) -I\left(U;S_{1} \right)  \right] \label{eq.4} 
\end{equation}
for all distributions:
\begin{equation}
p\left(y,x,u,s_{1}\right) =p\left(y\mid x,s_{1} \right)p\left(u,x\mid s_{1} \right)p\left(s_{1}\right) \label{eq.5}  
\end{equation}
where $ U $ is an auxiliary random variable.

\begin{figure}[!t]
\centering
\includegraphics[width=3.5in]{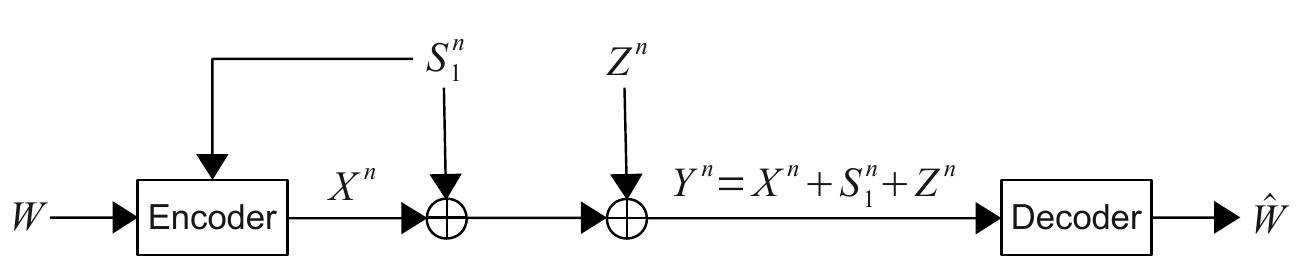}
\caption{Gaussian channel with additive interference known at the transmitter.} 
\label{figure3}
\end{figure}

\textit{Costa's "Writing on Dirty Paper":}
Costa \cite{costa} examined the Gaussian version of the channel with side information known at the transmitter (Fig.\ref{figure3}).

It is seen that the side information is considered as an additive interference at the receiver. Costa derived the capacity by using the result of Gel'fand-Pinsker theorem extended to random variables with continuous alphabets. The sketch of proof is as follows:
In Costa channel $ S_{1}^{n} $ is a sequence of Gaussian i.i.d. random variables with power $Q_{1}$. The transmitted sequence $ X^{n} $ is assumed to have the power constraint $ E\left\lbrace X^{2} \right\rbrace \leq P $. The output $ Y^{n}=X^{n}+S_{1}^{n}+Z^{n} $ where $Z^{n} $ is the sequence of white Gaussian noise with zero mean and power $ N $ ($ Z\sim \mathcal{N}\left(0,N \right)  $) and independent of both $ X $ and $ S_{1} $.\\
Costa established the capacity by obtaining a lower bound and an upper bound and proving the equality of these two bounds.
Although there is no definite condition on correlation between the channel input $X$ and the known interference $S_{1}$ in Costa channel, the achievable rate of $\frac{1}{2}\log\left(1+\frac{P}{N} \right)$ is obtained by taking $S_{1}$ and $X$ independent and the auxiliary random variable $U$ in (\ref{eq.5}) as $ U=\alpha\ S_{1}+X $. On the other hand, it can be shown that:
\begin{equation}
C\leq \max_{p\left (x\mid s_{1}\right )}\left[ I\left (X,Y\mid S_{1}\right )\right] \leq \frac{1}{2}\log\left( 1+\frac{P}{N}\right) \label{eq.6}
\end{equation}
so $\frac{1}{2}\log\left(1+\frac{P}{N} \right)$ is an upper bound for the capacity of channel and then the capacity of channel. What is surprising is that the capacity is independent of $S_{1}$, and that the capacity is equal to the capacity of channel when there is no interference $S_{1}$.

\section{A Capacity Theorem for Analyzing the Impact of Transmitter-Receiver Cognition on Channel Capacity}
\begin{figure}[!t]
\centering
\includegraphics[width=3.5in]{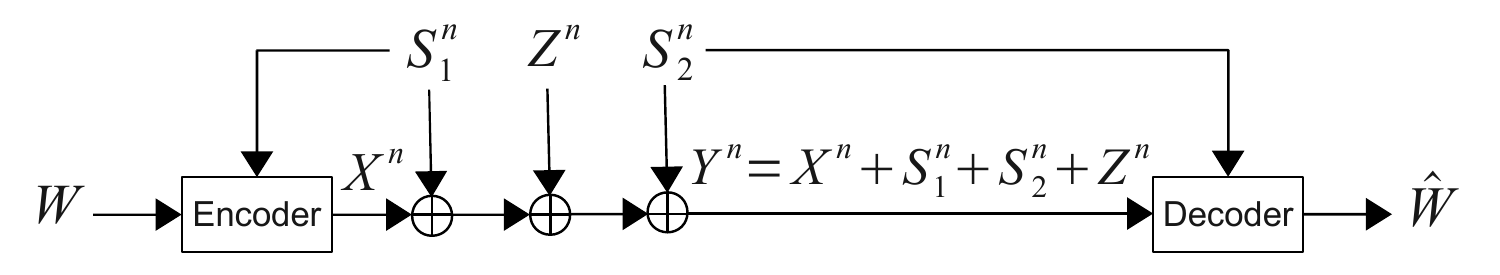}
\caption{Gaussian channel with correlated side information known at the transmitter and at the receiver. .}
\label{figure4}
\end{figure}

In this section we define and investigate a Gaussian channel in presence of two-sided information known non-causally at the transmitter and at the receiver. The side information at the transmitter and at the receiver is considered as additive interference at the receiver (Fig.\ref{figure4}).
In comparison with Costa channel, our channel has two major modifications: 1) In Costa channel there is no condition for the correlation between the channel input $X$ and the side information $S_{1}$. So $\frac{1}{2}\log \left (1+\frac{P}{N}\right )$ is the capacity of a channel in which the side information $S_{1}$ can be freely correlated to the channel input $X$; so this capacity can not be used for a channel with a specific correlation between $X$ and $S_{1}$. The correlation coefficient $\rho_{XS_{1}}$ between X and $S_{1}$ is specified in our channel. 2) We suppose that the Gaussian side information $S_{2}$ known at the receiver, exists and is correlated to the channel noise $Z$.

It is important to note that assuming the input random variable $X$ and $S_{1}$ correlated to each other with a specific correlation coefficient, does not impose any restriction on $X$'s own distribution and the distribution of $X$ is still free to choose.
\subsection*{\textbf{Definition of the Channel}}
 Consider the Gaussian channel depicted in Fig.\ref{figure4}. Our channel is defined with properties D.1-D.3 below:
\paragraph*{\textbf{D.1}}
$\left( S_{1}^{n},S_{2}^{n}\right)   $ are i.i.d.  sequences with zero mean and jointly Gaussian distributions.
\paragraph*{\textbf{D.2}}
Random variables $ \left(X,S_{1},S_{2} \right) $ have the covariance matrix $ \boldsymbol{K} $:
\begin{equation}
\boldsymbol{K}=
\begin{bmatrix}
\sigma_{X}^{2}&\sigma_{X}\sigma_{S_{1}}\rho_{XS_{1}}&\sigma_{X}\sigma_{S_{2}}\rho_{XS_{2}}\\
\sigma_{X}\sigma_{S_{1}}\rho_{XS_{1}}&\sigma_{S_{1}}^{2}&\sigma_{S_{1}}\sigma_{S_{2}}\rho_{S_{1}S_{2}}\\
\sigma_{X}\sigma_{S_{2}}\rho_{XS_{2}}&\sigma_{S_{1}}\sigma_{S_{2}}\rho_{S_{1}S_{2}}&\sigma_{S_{2}}^{2}
\end{bmatrix}. \label{eq.7}
\end{equation} We suppose that $S_{2}$ is independent of $X$ and $S_{1}$, so we have $\rho_{XS_{2}}=\rho_{S_{1}S_{2}}=0$. Moreover  $ X^{n} $ is assumed to have the constraint $ E\left\lbrace X^{2}\right\rbrace=\sigma_{X}^{2}\leq P  $. All values in $ \boldsymbol{K} $ except $ \sigma_{X} $, are fixed and must be considered as the definition of the channel. 
\paragraph*{\textbf{D.3}}
The output sequence  $ Y^{n}=X^{n}+S_{1}^{n}+S_{2}^{n}+Z^{n} $, where $ Z^{n} $ is the sequence of white Gaussian noise with zero mean and power $\sigma_{Z}^{2}= N $ ($ Z\sim \mathcal{N}\left( 0,N\right)  $) and independent of $ (X ,S_{1}) $ and \textit{dependent on} $S_{2}$ with $\rho_{S_{2}Z}$. For simplicity, we define: 
\begin{equation}
L_{2}\triangleq E\left\lbrace S_{2}Z\right\rbrace =\sigma_{S_{2}}\sigma_{Z}\rho_{S_{2}Z}.\label{eq.115} 
\end{equation} 
\paragraph*{\textbf{D.4}}
$ \left( X,U,S_{1},S_{2}\right)$ form the Markov Chain $ S_{2}\rightarrow S_{1}\rightarrow UX $. (We note that as mentioned earlier, this Markov chain (\ref{eq.3}) must be satisfied by all distributions $ p\left(y,x,u,s_{1},s_{2}\right)$ in Cover-Chiang capacity theorem and is physically acceptable).

It is readily seen that all distributions $ p\left(y,x,u,s_{1},s_{2}\right)$ specified with D.1-D.4 are in the form of (\ref{eq.2}) and hence we can use the extended version of Cover-Chiang theorem to random variables with continuous alphabets about the capacity of this channel.

Comparing our channel (defined with D.1-D.4) with Costa channel , a question may arise: (if we ignore $S_{2}$,) what is the relationship between capacities of these channels? To answer this question let us consider a subset of all distributions (channels)  $ p\left(y,x,u,s_{1}\right)$ (ignoring $S_{2}$) that satisfy D.1-D.4 and are similar but with different $\rho_{XS_{1}}$. Since Costa channel imposes no restriction on $\rho_{XS_{1}}$, these channels differ from the corresponding Costa channel on the restricted $\rho_{XS_{1}}$. It is clear that searching for the capacity of the Costa channel is led to the maximum capacity in this subset. So if $C_{D}$ be the capacity of the channel defined with D.1-D.4, and $C$ be the capacity of the Costa channel, we can write:
\begin{equation}
C=\max_{\rho_{XS_{1}},\rho_{S_{2}Z}=0}C_{D}.\label{eq.34+1} 
\end{equation}
 
We will show that the situation that $(X,S_{1},S_{2})$ are jointly Gaussian  and the auxiliary random variable $U$ is designed as linear combination of $X$ and $S_{1}$, is optimum and maximizes the transmitting rate. So we consider an important subset of the distributions $ p\left(y,x,u,s_{1},s_{2}\right)$ defined in D.1-D.4, as the set of all $ p^{\ast}\left(y,x,u,s_{1},s_{2}\right)$ that have the properties D.5 and D.6 below, in addition to D.1-D.4 (although the channel is defined only with D.1-D.4) :
\paragraph*{D.5}
Random variables $ \left( X,S_{1},S_{2}\right)$ are jointly Gaussian distributed. $ X $ is with zero mean and has the maximum power of $ P $ (so $ X\sim \mathcal{N}\left(0,P \right)$).
Naming the covariance matrix in this special case as $ \boldsymbol{K^{\ast}} $, for simplicity, by defining $ A_{1}\triangleq E\left\lbrace XS_{1}\right\rbrace $ , $Q_{1}\triangleq \sigma_{S_{1}}^{2}$ and $Q_{2}\triangleq \sigma_{S_{2}}^2$, we rewrite:
\begin{equation}
\boldsymbol{K^{\ast}}=\begin{bmatrix}
P&A_{1}&0\\
A_{1}&Q_{1}&0\\
0&0&Q_{2}
\end{bmatrix}\label{eq.35}
\end{equation}
\paragraph*{D.6}
Following Costa, we consider $ U $ in the form of linear combination of $ X $ and $ S_{1} $ as $U=\alpha S_{1}+X$.

For summarizing expressions, we define two following symbols:
\setlength{\arraycolsep}{0.0em}
\begin{eqnarray}
d_{Q_{2}}  &\triangleq & PQ_{1}-A_{1}^{2}= \sigma_{X}^{2}\sigma_{S_{1}}^{2}\left (1-\rho_{XS_{1}}^{2}\right ) \label{eq.37} \\
d_{PQ_{1}}&\triangleq & Q_{2}N-L_{2}^{2}= \sigma_{S_{2}}^{2}\sigma_{Z}^{2}\left (1-\rho_{S_{2}Z}^{2}\right ). \label{eq.102} 
\end{eqnarray}
\setlength{\arraycolsep}{5pt}
\subsection*{\textbf{Capacity of the Channel}}
\paragraph*{Theorem 1}
The Gaussian channel defined with properties D.1-D.4 has the capacity
\begin{equation}
C_{D}=\dfrac{1}{2}\log\left (1+\dfrac{P\left (1-\rho_{XS_{1}}^{2}\right )}{N\left (1-\rho_{S_{2}Z}^{2}\right )}\right )\label{eq.116} 
\end{equation}

\textit{Corollary 1:} As mentioned earlier, by (\ref{eq.34+1}) we can obtain Costa capacity by assuming $\rho_{S_{2}Z}=0$ and maximizing $C_{D}$  with $\rho_{XS_{1}}=0$.

\textit{Corollary 2:} It is seen that if the side information $S_{2}$ is independent of the channel noise $Z$ (and so $\rho_{S_{2}Z}=0$), the capacity of the channel is equal to the capacity when there is no interference $S_{2}$. In other words, in this case the receiver can subtract the known $S_{2}^{n}$ from the received $Y^{n}$ without losing any worthy information. But when the state information $S_{2}$ is correlated with additive noise $Z$, $S_{2}$ is containing worthy information that increases the capacity, and hence subtracting $S_{2}$ is a wrong decoding strategy. 

\textit{Corollary 3:} It is seen that while, as intuitively expected, correlation between $S_{2}$ and $Z$ increases the capacity, the correlation between $X$ and $S_{1}$ decreases it.\\
\paragraph*{Proof of Theorem 1}
To prove the theorem, we first show that $C_{D}$ (\ref{eq.116}) is a lower bound for the capacity of the channel, then we show that $C_{D}$ is an upper bound for the capacity too, so $C_{D} $ is the capacity of the channel.
\subsubsection*{Achievability part of the proof} 
we use the extended version of Cover-Chiang capacity (\ref{eq.1}) to obtain a lower bound for the capacity of the channel: For all distributions $ p\left (y,x,u,s_{1},s_{2}\right ) $ (with properties D.1-D.4) and its subset $ p^{\ast}\left (y,x,u,s_{1},s_{2}\right ) $ (defined with properties D.1-D.6), we can write:
\setlength{\arraycolsep}{0.0em}
\begin{eqnarray}
C&=&\max_{p\left (u,x\mid s_{1}\right )}\left[ I\left (U;Y,S_{2}\right )-I\left (U;S_{1}\right )\right ]\label{eq.49}\\
&\geq &\max_{p^{\ast}\left (u\mid x,s_{1}\right )p^{\ast}\left (x\mid s_{1}\right )}\left[ I\left (U;Y,S_{2}\right )-I\left (U;S_{1}\right )\right ]\label{eq.51}\\
&=&\max_{\alpha}\left[ I\left (U;Y,S_{2}\right )-I\left (U;S_{1}\right )\right ]\label{eq.52}\\
&\triangleq &\max_{\alpha}R_{D}\left (\alpha\right )=R_{D}\left (\alpha^{\ast}\right ). \label{eq.53}
\end{eqnarray}
\setlength{\arraycolsep}{5pt}
So $R_{D}\left (\alpha^{\ast}\right )$ is a lower bound for the capacity of the channel. To compute $R_{D}\left (\alpha\right )$ we write (details of computations are omitted for the brevity):
\setlength{\arraycolsep}{0.0em}
\begin{eqnarray}
I\left (U;Y,S_{2}\right )&=&H\left (U\right )+H\left (Y,S_{2}\right )-H\left (U,Y,S_{2}\right ),\label{eq.99}\\
I\left (U;S_{1}\right )&=&H\left (U\right )+H\left (S_{1}\right )-H\left (U,S_{1}\right ),\label{eq.100}
\end{eqnarray}
where
\begin{eqnarray}
&&H\left (Y,S_{2}\right)=\dfrac{1}{2}\log\left (\left (2\pi e\right )^{2}\det\left (cov\left (Y,S_{2}\right )\right )\right )\label{eq.101}\\
&& \quad=\dfrac{1}{2}\log\bigg( (2\pi e)^{2}\big(Q_{2}(P+Q_{1}+2A_{1})+d_{PQ_{1}}\big)\bigg),\nonumber
\end{eqnarray}
\begin{eqnarray}
H\left (U,Y,S_{2}\right )=\dfrac{1}{2}\log \bigg((2\pi e)^{3} &\big[& d_{PQ_{1}}(\alpha ^{2}Q_{1}+2\alpha A_{1}+P)\nonumber\\
&+&(\alpha -1)^{2}Q_{2}d_{Q_{2}}\big]\bigg),\label{eq.103}  
\end{eqnarray}
\begin{eqnarray}
H\left (S_{1}\right )&=&\dfrac{1}{2}\log\left (\left (2\pi e\right )Q_{1}\right ),\label{eq.104}\\ 
H\left (U,S_{1}\right )&=&\dfrac{1}{2}\log \left (\left (2\pi e\right )^{2}d_{Q_{2}}\right ).\label{eq.105}
\end{eqnarray}
\setlength{\arraycolsep}{5pt}
Substituting (\ref{eq.101})-(\ref{eq.105}) in (\ref{eq.99}) and (\ref{eq.100}), we obtain:
\begin{small}
\setlength{\arraycolsep}{0.0em}
\begin{eqnarray}
&&R_{D}\left(\alpha\right)=\label{eq.108}\\
&&\dfrac{1}{2}\log\left( \dfrac{d_{Q_{2} }\left(Q_{2}\left (P+Q_{1}+2A_{1}\right )+d_{PQ_{1}}\right )}{Q_{1}\left(\left(\alpha -1\right)^{2}Q_{2}d_{Q_{2}}+d_{PQ_{1}}\left(\alpha^{2}Q_{1}+2\alpha A_{1}+P\right)   \right) }\right) \nonumber 
\end{eqnarray}
\setlength{\arraycolsep}{5pt} 
\end{small}
and after maximizing it over $ \alpha $, we conclude:
\begin{equation}
\alpha^{\ast}=\dfrac{Q_{2}d_{Q_{2}}-A_{1}d_{PQ_{1}}}{Q_{2}d_{Q_{2}}-Q_{1}d_{PQ_{1}}}.\label{eq.106}
\end{equation}
Now, if we compute $R_{D}\left (\alpha^{\ast}\right )$ by putting (\ref{eq.106}) into (\ref{eq.108}) and then rewrite the resulted expression in terms of $\sigma_{X}$, $\sigma_{S_{1}}$, $\sigma_{S_{2}}$, $\rho_{XS_{1}}$, $\rho_{S_{2}Z}$ by (\ref{eq.115}) and (\ref{eq.35})-(\ref{eq.102}) we finally conclude:
\begin{equation}
R_{D}\left (\alpha^{\ast}\right )=\dfrac{1}{2}\log\left (1+\dfrac{P\left (1-\rho_{XS_{1}}^{2}\right )}{N\left (1-\rho_{S_{2}Z}^{2}\right )}\right ) \label{eq.120} 
\end{equation}

\subsubsection*{Converse part of the proof}
For all distributions $ p\left(y,x,u,s_{1},s_{2}\right)$ defined with properties D.1-D.4, we have:
\setlength{\arraycolsep}{0.0em}
\begin{eqnarray}
I\left( U;Y,S_{2}\right)-I\left( U;S_{1}\right)&=&-H\left( U\mid Y,S_{2}\right) + H\left(U\mid S_{1} \right)\nonumber  \\
&\leq &I\left (X;Y\mid S_{1},S_{2}\right )\label{eq.77}
\end{eqnarray}
\setlength{\arraycolsep}{5pt}
where (\ref{eq.77}) follows from  Markov  chains  $ S_{2}\rightarrow S_{1}\rightarrow UX $ and  $ U\rightarrow XS_{1}S_{2}\rightarrow Y $, which are true for all distributions defined with properties D.1-D.4. Now from (\ref{eq.1}) and (\ref{eq.77}) we can write:
\setlength{\arraycolsep}{0.0em}
\begin{eqnarray}
C&=&\max_{p\left ( u,x\mid s_{1}\right ) } \left [ I\left ( U;Y,S_{2}\right ) -I\left ( U;S_{1}\right )\right ]\label{eq.78}\\
&\leq & \max_{p\left ( x\mid s_{1}\right )}\left [I\left ( X;Y\mid S_{1},S_{2}\right ) \right ]\triangleq I^{\ast}\left (X;Y\mid S_{1},S_{2}\right ),\label{eq.79}
\end{eqnarray}
\setlength{\arraycolsep}{5pt}
hence $I^{\ast}\left (X;Y\mid S_{1},S_{2}\right )$ is an upper bound for the capacity of the channel. For computing it we write:
\setlength{\arraycolsep}{0.0em}
\begin{eqnarray}
&&I\left ( X;Y\mid S_{1},S_{2}\right )\nonumber\\
&&=H\left (\left (X+Z\right ),S_{1},S_{2}\right )-H\left (S_{1},S_{2}\right )-H\left (Z\mid S_{2}\right ).\label{eq.83}
\end{eqnarray}
\setlength{\arraycolsep}{5pt}
So when (\ref{eq.79}) reaches to its maximum, $ \left (X,S_{1},S_{2}\right )$ are jointly Gaussian and $ X $ has its maximum power of $ P $ and it means that $I^{\ast}\left (X;Y\mid S_{1},S_{2}\right ) $ is the value of (\ref{eq.83}) which is computed for distributions $ p^{\ast}\left (y,x,s_{1},s_{2}\right ) $ defined with properties D.1-D.6. After computing we have:

\begin{small}
\setlength{\arraycolsep}{0.0em}
\begin{eqnarray}
H\left (\left (X+Z\right ),S_{1},S_{2}\right )&=&\dfrac{1}{2}\log\left( \left( 2\pi e\right) ^{3}\left(Q_{2}d_{Q_{2}}+Q_{1}d_{PQ_{1}}\right) \right) \label{eq.112} \\
H\left (S_{1},S_{2}\right )&=&\dfrac{1}{2}\log\left (\left (2\pi e\right )^{2}Q_{1}Q_{2}\right )\label{eq.113} \\
H\left (Z,S_{2}\right )&=&\dfrac{1}{2}\log\left (\left (2\pi e\right )^{2}d_{PQ_{1}}\right )\label{eq.114} 
\setlength{\arraycolsep}{0.0em}
\end{eqnarray}
\setlength{\arraycolsep}{5pt}
\end{small}
so we obtain from (\ref{eq.83})-(\ref{eq.114}):
\begin{eqnarray}
I^{\ast}\left (X;Y\mid S_{1},S_{2}\right )&=&\dfrac{1}{2}\log \left (\dfrac{Q_{2}d_{Q_{2}}+Q_{1}d_{PQ_{1}}}{Q_{1}d_{PQ_{1}}}\right )\label{eq.110} \\
&=&\dfrac{1}{2}\log\left( 1+\dfrac{P\left( 1-\rho_{XS_{1}}^{2}\right)}{N\left( 1-\rho_{S_{2}Z}^{2}\right)} \right).  \label{eq.111}
\end{eqnarray}
\setlength{\arraycolsep}{5pt}
where (\ref{eq.111}) follows by rewriting (\ref{eq.110}) in terms of $\sigma_{X}$, $\sigma_{S_{1}}$, $\sigma_{S_{2}}$, $\rho_{XS_{1}}$, $\rho_{S_{2}Z}$ by (\ref{eq.115}) and (\ref{eq.35})-(\ref{eq.102}).

From (\ref{eq.120}) and (\ref{eq.111}), we conclude that $C_{D}$ (\ref{eq.116}) is the capacity of the channel.\qed

\section{Numerical Results}
\begin{figure}[!t]
\centering
\includegraphics[width=2.2in]{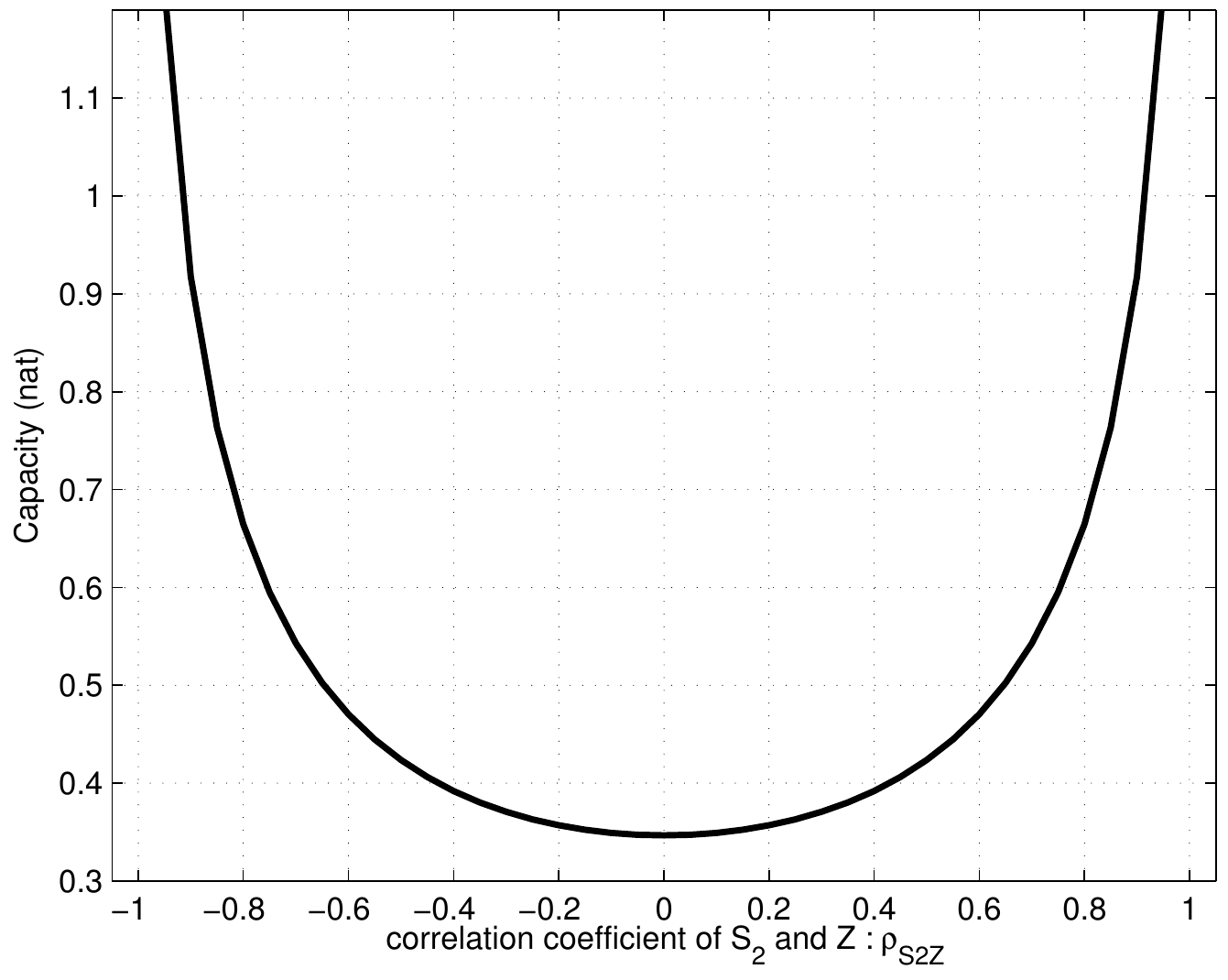}
\caption{The impact of the correlation of state information $S_{2}$ and noise $Z$ on the capacity of the channel. Figure plotted with $\rho_{XS_{1}}=0$.}
\label{figure6}
\end{figure}

Fig.\ref{figure6} illustrates the impact of the correlation of $S_{2}$ and the channel noise $Z$ on the channel capacity. Figure plotted for independent $X$ and $S_{1}$ (so $\rho_{XS_{1}}=0 $). It is seen that the more $S_{2}$ depends on the noise $Z$ , the greater capacity of channel is. On the condition of full dependency $\rho_{S_{2}Z}=\pm 1$ the capacity of channel is infinite.

\section{Conclusion}
We investigated the Gaussian channel in the presence of two-sided state information with dependency on the input and the channel noise. Having established a capacity theorem for the channel, we illustrated the impact of the receiver cognition (the correlation between the channel noise and state information known at the receiver) and the correlation between the input and the side information known at the transmitter, on the capacity of the channel.

\bibliographystyle{IEEEtran}	
\bibliography{Capacitybiblio}		

\end{document}